\documentclass[pra,10pt,twocolumn,showpacs]{revtex4}
\usepackage[dvips]{epsfig}
\usepackage{float}
\usepackage{upgreek}
\usepackage{amsfonts}
\usepackage{amsmath}
\usepackage{amssymb}
\usepackage{enumerate}
\usepackage{pslatex}
\usepackage{hhline}
\usepackage{threeparttable}
\usepackage{supertabular}
\usepackage{multirow}
\usepackage{tabularx}

\newcommand{\vrr}{\mathbf{r}}

\begin{document}

\bibliographystyle{apsrev}

\title{Chip-based microcavities coupled to NV centers in single crystal diamond}

\author{Paul E. Barclay, Kai-Mei C. Fu, Charles Santori and Raymond G. Beausoleil}
\address{Hewlett-Packard Laboratories, 1501 Page Mill Road, Palo Alto CA 94304}
\email{paul.barclay@hp.com}

\begin{abstract}
Optical coupling of nitrogen vacancy centers in single-crystal diamond to an on-chip microcavity is demonstrated. The microcavity is fabricated
from a hybrid gallium phosphide and diamond material system, and supports whispering gallery mode resonances with spectrometer resolution
limited $Q > 25000$.
\end{abstract}


\maketitle

\noindent An outstanding challenge in creating solid state cavity QED systems \cite{ref:kimble1998sis} useful for quantum information processing
is efficient optical coupling to high quality spin qubits \cite{ref:benjamin2009pfm}.  The nitrogen-vacancy (NV) center in diamond is a
promising candidate qubit, as it allows optical readout of long-lived single electron \cite{ref:jelezko2004oco} and nuclear
\cite{ref:jelezko2004ocs, ref:gurudevdutt2007qrb, ref:childress2006cdc} spins. Using coherent optical control \cite{ref:santori2006cpt}, it may
be possible to generate single photons \cite{ref:kurtsiefer2000sss,ref:beveratos2002rts} entangled with a single NV spin
\cite{ref:cabrillo1999ces, ref:benjamin2009pfm}. Chip-based nanophotonic devices such as microcavities can play an important role in these
experiments, providing efficient optical coupling through Purcell-enhanced qubit-photon interactions, and scalable optical integration of
multiple qubits. Studies of optical coupling between NVs in diamond nanocrystals and high-$Q$ dielectric microcavities \cite{ref:park2006cqd,
ref:barclay2008cie, ref:schietinger2008obo} have been limited by the relatively poor nanocrystal NV optical properties, and NVs have not been
observed in nanocrystalline diamond microcavities \cite{ref:wang2007owg, ref:wang2007fct}. The alternative, nanophotonic coupling to high
quality NVs in single crystal diamond, is challenging, owing to the difficulty in realizing waveguiding structures from bulk diamond.

Here we report the demonstration of high-$Q$ microcavities fabricated in single crystal diamond by integrating a patterned high-index
waveguiding layer on a diamond substrate, from which photons couple evanescently to NVs close to the diamond surface. This approach was first
studied in Ref.\ \cite{ref:fu2008cnv}, where waveguides patterned in a gallium phosphide film (GaP, $n_\text{GaP}=3.25$) were optically coupled
to NVs in a diamond substrate ($n_\text{dia}=2.42$). Evanescent coupling was also used in Ref.\ \cite{ref:larsson2009com}, where an SiO$_2$
microsphere cavity was coupled to NVs in a diamond pillar \cite{ref:babinec2009bsp}. In this letter, we measure coupling between microcavities
\emph{on-chip} and NV centers in a diamond substrate. The monolithic, chip-based nature of these devices is amenable to fabrication of more
complicated nanophotonic structures such as photonic crystal nanocavities and waveguides \cite{ref:barclay2009hpc}.

\begin{figure}[hbt]
\begin{center}
  \epsfig{figure=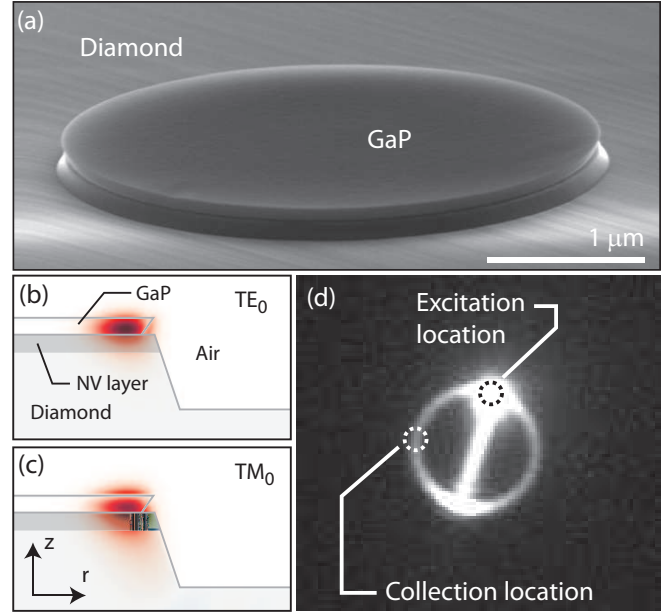, width=1\linewidth}
  \caption{(a) Scanning electron microscope (SEM) image of a hybrid GaP-diamond microdisk. (b,c) FDTD simulated field profiles
  ($E_r(r,z)$ and $E_z(r,z)$, respectively), of the TE$_0^m$ and TM$_0^m$ modes.  (d) Widefield CCD image of photoluminescence from a
  hybrid microdisk.
  }\label{fig:SEM}
\end{center}
\end{figure}

The devices studied here consist of GaP microdisks supported by a single crystal diamond substrate; an SEM image of a typical device is shown in
Fig.\ \ref{fig:SEM}(a). In order to reduce radiation loss into leaky substrate modes, the sidewalls of the GaP structure are extended into the
underlying diamond. These hybrid microdisks \cite{ref:tien2009hii} support high-$Q$ whispering gallery modes which interact evanescently with
the diamond substrate. Figures \ref{fig:SEM}(b) and \ref{fig:SEM}(c) show cross-sections of the electric field of the lowest order whispering
gallery modes with dominantly radial (TE) and vertical (TM) electric field polarization, calculated using three dimensional finite difference
time domain (FDTD) simulations with $e^{im\phi}$ azimuthal field variation \cite{ref:meep}.  These modes are labeled TE$^m_0$ and TM$^m_0$,
indicating a single maximum in the radial dimension, and azimuthal quantum number $m$.

\begin{table}[tb]
\begin{tabular}{l c c r}
Sample & Supplier & GaP thickness ($t$) & N$^+$ implantation \\  \hline\hline
HPHT & Sumitomo  & 250 nm & 50 keV, $2\times10^{13}$/cm$^{2}$ \\
CVD & Element 6 & 130 nm & 10 keV, $2\times10^{13}$/cm$^{2}$ \\
\end{tabular}
\caption{Hybrid microdisk sample properties.}\label{tab:sample}
\end{table}

\begin{figure}[t]
\begin{center}
  \epsfig{figure=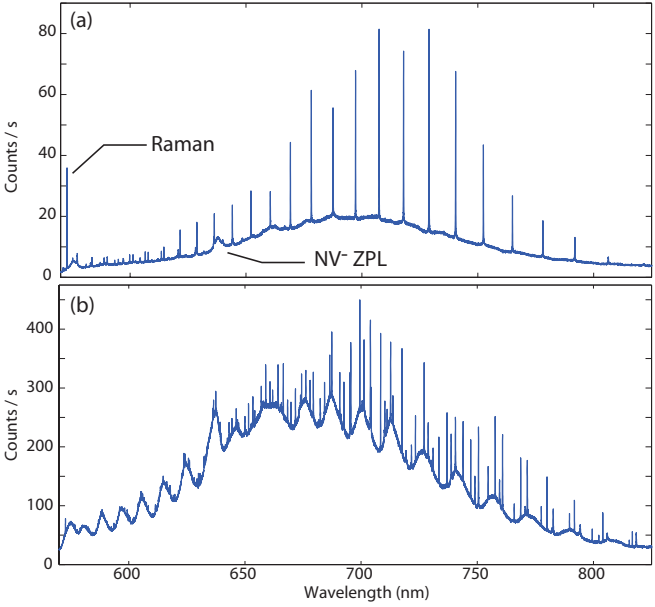, width=1\linewidth}
  \caption{PL spectra of GaP-diamond microdisks excited with a 532 nm source.  Microdisk thickness and diameter
  $d \sim \left[4.6,4.5\right]\mu$m, $t \sim \left[130, 250\right]$nm for the devices measured in (a,b), respectively. Diamond etch
  depth $h \sim 600$ nm. Excitation power $\sim 2.5$ mW}\label{fig:data}
\end{center}
\end{figure}

Two hybrid diamond-GaP microdisk samples were fabricated, as summarized in Table \ref{tab:sample}. NVs were created within $\sim 200$ nm of the
diamond sample surface using N$^+$ ion implantation, followed by a one hour $950~^\text{o}$C hydrogen-argon anneal \cite{ref:davies1992vrc,
ref:meijer2005gsc, ref:santori2008vdn}.  The diamond samples were then cleaned in a H$_2$SO$_4$:KNO$_3$ (20mL:1g, $\sim 240~^\text{o}$C)
solution. For fabrication of the GaP microdisks, we used a GaP layer (thickness $t$) grown expitaxially on an AlGaP sacrificial layer on a GaP
substrate. Electron beam lithography followed by Ar/BCl$_3$/Cl$_2$ ECR-RIE anisotropic plasma etching and HF wet etching (7\% 1:1 HF:H$_2$O) was
used to define GaP microdisks supported by AlGaP pillars on a GaP substrate \cite{ref:srinivasan2005oll}. To transfer the microdisks to the
diamond sample, a drop of HF was placed on the diamond sample surface and held by surface tension. The GaP microdisk sample was placed with its
top surface facing down on the HF covered diamond surface, and held in place for 5 minutes. The HF completely removes the AlGaP pillars, and the
GaP microdisks either fall onto the diamond surface, or attach to the GaP substrate. The GaP substrate was detached from the diamond in a H$_2$O
bath. After an N$_2$ drying step, a large fraction of the GaP microdisks remain attached to the diamond top surface. An anisotropic O$_2$ plasma
ICP-RIE etch was then used to selectively remove up to $600$ nm from the diamond surface not masked by the GaP structures. As discussed below
and in Ref.\ \cite{ref:barclay2009hpc}, extending microdisk sidewalls into the diamond substrate allows smaller structures to be realized for a
chosen minimum radiation loss limited $Q$. Although the relative device positions are not fixed in this work, connected patterns such as
integrated photonic crystal devices \cite{ref:barclay2009hpc} may be compatible with this process.

\begin{figure}[t]
\begin{center}
  \epsfig{figure=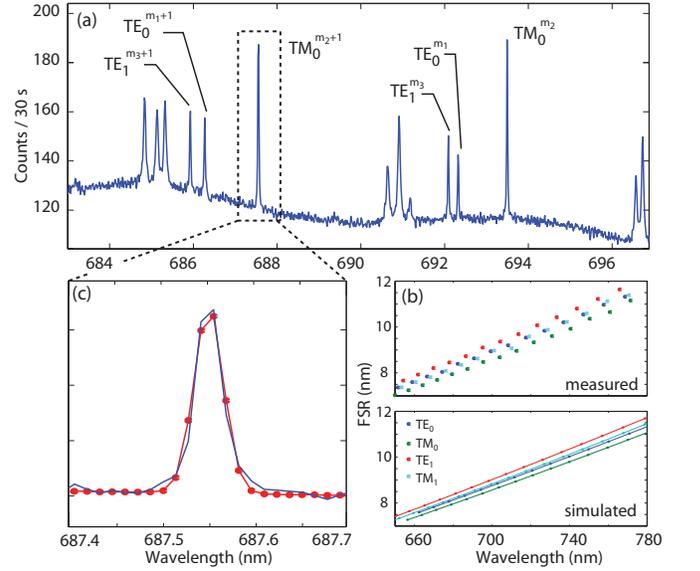, width=1\linewidth}
  \caption{(a) PL spectra of a $\left[t,d\right] = \left[0.25,6.5\right]\mu$m microdisk.  From FDTD simulations we
  estimate that $\left[m_1,m_2,m_3\right] =  \left[83,82,77\right]$. (b) Measured (top) and FDTD calculated (bottom) FSR dispersion of the
  four highest-$Q$ sets of modes in Fig.\ \ref{fig:data}(b). GaP refractive index dispersion \cite{ref:aspnes1983dfo} was included in FDTD simulations.
  (c) System response limited lineshape of the TM$_0^m$ resonance. Fit of a Lorentzian convolved with the pixelized Gaussian system response shown in red.}
  \label{fig:q_data}
\end{center}
\end{figure}

Optical coupling of NV photoluminescence (PL) into microdisk cavity modes was studied using a confocal microscope to optically excite NVs near
the microdisk edge, as shown in Fig.\ \ref{fig:SEM}(d). A green excitation source (532 nm excitation laser) was focused to a $\sim 0.5$ $\mu$m
diameter spot using an NA = 0.6 microscope objective.  A dichroic mirror and long wavelength pass filter (cut-on $\sim 540$ nm) were used to
spectrally separate reflected laser light from the PL.  The objective was used to collect PL from a region of the microdisk edge at an azimuthal
coordinate rotated $\sim 90^\text{o}$  from the excitation spot.  This spatial filtering was accomplished by focusing the collected PL through a
50 $\mu$m diameter pinhole at 37.5X magnification.

Figures \ref{fig:data}(a) and (b) show typical PL spectra for hybrid microdisks from the CVD and HPHT samples, respectively.  In both spectra, a
broad background corresponding to emission from NV centers is visible, as evidenced by the peak at $\lambda \sim 637$ nm from the NV$^-$ zero
phonon line (ZPL), and a global maximum near 680 nm from phonon assisted NV$^-$ emission.  Also visible is bulk diamond Raman emission at 572
nm. Superimposed upon this background are sharp peaks resulting from NV emission coupled into the microdisk modes, a fraction of which is then
scattered or radiated into the microscope objective.  The CVD device measured in Fig.\ \ref{fig:data}(a) has a GaP thickness of 130 nm, and for
$\lambda > 650$ nm only supports high-$Q$  TE$^m_0$ whispering gallery modes. The HPHT device measured in Fig.\ \ref{fig:data}(b) has $t \sim
250$ nm, and in addition to TE$^m_0$ modes, supports high-$Q$ TM$^m_0$, TE$^m_1$ and TM$^m_1$ modes. Also visible in the spectra are low-$Q$
oscillations resulting from Fabry-P\'erot-like transverse microcavity modes, whose field distribution along the microdisk diameter ($d$) is
evident in Fig.\ \ref{fig:SEM}(d). The order of magnitude difference in measured PL intensity between Figs.\ \ref{fig:data}(a) and
\ref{fig:data}(b) is a result of the higher implanted NV density of the relatively nitrogen rich HPHT sample compared to the CVD sample.

A spectrum illustrating the detailed mode structure of an HPHT hybrid microdisk is shown in Fig.\ \ref{fig:q_data}(a). Regularly spaced sets of
resonances with incrementing $m$-numbers are evident.  The polarization and radial mode labels for a given resonance can be determined by
measuring its free spectral range (FSR) as a function of wavelength, and comparing with FDTD simulated values. The results are shown in Fig.\
\ref{fig:q_data}(b) for four families of resonances in Fig.\ \ref{fig:data}(b); also shown are FDTD simulated FSR dispersion of the TE$_0^m$,
TM$_0^m$, TE$_1^m$ and TM$_1^m$ modes.


\begin{figure}[t]
\begin{center}
  \epsfig{figure=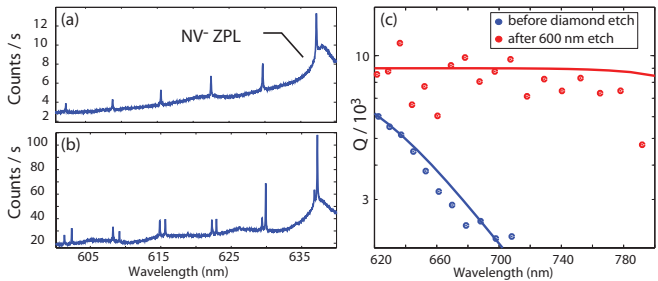, width=1\linewidth}
  \caption{Effect of diamond etching on microdisk mode spectra ($\left[t,d\right]=\left[0.130,4.5\right]\mu$m). Photoluminescence spectra
  (a) before and (b) after 250 nm vertical diamond etching.  Excitation power in (a,b) $\sim \left[0.25,2.5\right]$mW. (c)
  Wavelength dependence of $Q$ for the TE-0 mode before and after 600 nm diamond etching.  Simulated values for $Q$ shown by solid lines,
  assuming $1/Q = 1/Q_\text{rad} + 1/Q_i$ where radiation loss limited $Q_\text{rad}(\lambda)$ is calculated using FDTD, and $Q_i = 9000$.
   }\label{fig:etching_effects}
\end{center}
\end{figure}

The largest peak in Fig.\ \ref{fig:q_data}(a) is associated with a TM$_0^m$ resonance, and has a linewidth which is not measurably broader than
the spectrometer resolution.  A fit to this resonance, derived from a Lorentzian convolved with a Gaussian approximation for the system response
function (calibrated using spectral lines from a Hg lamp), is shown in Fig.\ \ref{fig:q_data}(c). The best-fit linewidth of the Lorentzian was
smaller than the error in the measured system response, placing a lower limit of $Q > 2.5\times10^4$.  Two other sets of resonances in Fig.\
\ref{fig:q_data}(a), associated with the TE$_0^m$ and TE$_1^m$ microdisk modes, also have linewidths not measurably larger than the system
resolution. Further studies, for example using laser spectroscopy, will be necessary to accurately measure $Q$ for these modes.  For this
device, the TM$_0^m$ mode interacts most strongly with NVs near the diamond surface, as it has both the smallest mode volume, $\overline{V}$,
and the largest relative field strength in the diamond substrate, $\eta$ \endnote{$\overline{V} = (\lambda/n_\text{GaP})^{-3}\int
n^2(\vrr)|E(\vrr)|^2 d\vrr /(n^2(\vrr_o)|E(\vrr_o)|^2)$, and $\eta = E(\vrr|n(\vrr) = n_\text{dia})|_\text{max} / E(\vrr_o)$ where $n^2|E|^2$ is
maximized at  $\vrr_o$.}. From FDTD simulations, $\left[\overline{V},\eta\right] = \left[43,0.48\right]$ for a TM$_0^{89}$ standing wave mode
supported by this device near $\lambda = 637$ nm.

Etching the diamond to extend the microdisk sidewalls into the substrate permits smaller $\overline{V}$ and larger $\eta$ devices to be realized
for a given radiation loss limited $Q$ \cite{ref:barclay2009hpc}. Figures \ref{fig:etching_effects}(a) and \ref{fig:etching_effects}(b) show
spectra for a $\left[d,t\right] = \left[4.5,0.13\right]\mu$m hybrid microdisk before and after the diamond etch step. Note the emergence of a
second family of previously low-$Q$ resonances (the TE$_1^m$ modes, for this device) after etching. Prior to etching the diamond, the TE$_0^m$
mode $Q$ decreases exponentially for $\lambda > 620$ nm due to radiation loss. After etching, the radiation loss cutoff is increased above $780$
nm. This is consistent with FDTD simulated $Q$ dispersion, as shown in Fig.\ \ref{fig:etching_effects}(c), assuming that the device has an
intrinsic $Q_i\sim 9000$. At $\lambda\sim 637$ nm, $\left[\overline{V},\eta\right]= [18,0.57]$ for the TE$_0^{56}$ mode of this device. The
relatively low $Q_i$, compared with that of the larger $t$ device in Fig.\ \ref{fig:q_data}, is a result of a lower GaP surface quality and
increased surface-field interaction.  The etched sidewalls have a 3 nm RMS roughness (80 nm correlation length) due to sub-optimal GaP etching
and lithography. It is estimated that this effect will limit the $Q<1.7\times10^4$ \cite{ref:borselli2005brs}. In addition, the GaP underside
has roughness from imperfect electron-beam resist removal. These imperfections can be improved in future devices.

These results provide a proof-of-principal demonstration of on-chip nanophotonic devices for efficient NV optical coupling. The relevant
cavity-QED parameters \cite{ref:kimble1998sis} for the $\left[d,t\right]=\left[4.5,0.13\right]\mu$m microdisks are
$\left[g_\text{ZPL},\kappa,\gamma,\gamma_\text{ZPL}\right]/2\pi = \left[0.30,26,0.013,0.0004\right]$GHz. Here, $g_\text{ZPL}$ is the coupling
strength \cite{ref:barclay2009hpc} between a single microdisk photon and the NV$^-$ ZPL, for an NV$^-$ optimally oriented and positioned at the
diamond surface, $\kappa=\omega/2Q$ is the microdisk photon decay rate, and $\gamma$ and $\gamma_\text{ZPL}$ are the \emph{total} and \emph{ZPL}
spontaneous emission rates of an NV$^-$. At low-temperature, the ZPL spontaneous emission rate of an optimally positioned NV$^-$ into a
microcavity mode is predicted to be enhanced by $F_\text{ZPL}\sim 17$. This corresponds to coupling $\beta \sim 34\%$ of the \emph{total} NV$^-$
spontaneous emission (e.g., including phonon sideband emission) into the microcavity. If the $Q$ of this device can be increased to $2.5 \times
10^4$, $F_\text{ZPL} \sim 47$ should be possible.  Similarly, a photonic crystal nanocavity \cite{ref:barclay2009hpc} fabricated from this
material system, with similar $Q$, will allow $\beta \to 1$, and provide a platform for on-chip integration of multiple devices required by
applications in quantum information processing.

\acknowledgements{PB thanks Oskar Painter for access to fabrication tools used in initial fabrication process development.  This work is
supported by the DARPA QUEST program.}


\end{document}